\documentclass[aps,preprint,showpacs]{revtex4}
\usepackage{graphicx}
\usepackage{dcolumn}
\begin{document}

\title{ Hyperon production in near threshold nucleon-nucleon collisions }
\author{Radhey Shyam}
\affiliation{$^1$Saha Institute of Nuclear Physics, Kolkata, India\\
}

\date{\today}

\begin{abstract}

We study the mechanism of the associated $\Lambda$-kaon and
$\Sigma$-kaon production in nucleon-nucleon collisions over an extended
range of near threshold beam energies within an effective Lagrangian
model, to understand the new data on $pp \to p\Lambda K^+$ and
$pp \to p\Sigma^0K^+$ reactions published recently
by the COSY-11 collaboration. In this theory, the hyperon production
proceeds via the excitation of $N^*(1650)$, $N^*(1710)$, and $N^*(1720)$
baryonic resonances. Interplay of the relative contributions of various
resonances to the cross sections, is discussed as a function of the beam
energy over a larger near threshold energy domain. Predictions of our model
are given for the total cross sections of $pp \to p\Sigma^+K^0$,
$pp \to n\Sigma^+K^+$, and $pn \to n\Lambda K^+$ reactions.
                                                                                
\end{abstract}
\maketitle

Dedicated experiments performed recently at the Cooler Synchrotron 
(COSY) facility at the Forschungszentrum, J\"ulich, have led to the
accumulation of high quality data on associated hyperon production
in close-to-threshold proton-proton ($pp$) collisions
\cite{bal98,sew99,kow04,bil98,abd04}. Strangeness production reactions
are expected to provide information on the manifestation of quantum
chromodynamics (QCD) in the non-perturbative regime of energies larger
than those of the low energy pion physics where the low energy theorem
and partial conservation of axial current (PCAC) constraints provide a
useful insight into the relevant physics~\cite{eric88}. The strangeness
quantum number introduced by this reaction leads to new degrees of
freedom in this domain which are likely to probe the admixture
of $\bar{s}s$ quark pairs in the nucleon wave function~\cite{albe96}.
At the near threshold beam energies, the final state interaction (FSI)
effects among the outgoing particles are significant. Therefore, the new
set of data are expected to probe also the hyperon-nucleon and
hyperon-strange meson interactions (see, e.g., Refs.~\cite{mos02,kel00,gas05}).

A very interesting result of the studies performed by the COSY-11
collaboration is that the ratio ($R$) of the total cross sections for
the $pp \to p\Lambda K^+$ and $pp \to p\Sigma^0 K^+$ reactions
at the same excess energy ($\epsilon$) [defined as $\epsilon =
\sqrt{s}-m_p-m_Y-m_K$, with $m_p$, $m_Y$, and $m_K$ being the masses
of proton, hyperon (Y), and kaon, respectively, and $\sqrt{s}$ the invariant
mass of the collision], is about $28^{+6}_{-9}$ for $\epsilon$ $<$ 13
MeV~\cite{sew99}. This result is very intriguing because at higher beam
energies this ratio is only about 2.5~\cite{lan88}.

Assuming that the hyperon production
proceeds solely due to the kaon($K$)-exchange mechanism and that the
final state interaction (FSI) effects among the outgoing particles are
absent, $R$ is given essentially by the ratio of the squares of coupling
constants at the vertices from which the $K^+$ meson emerges
($g_{N\Lambda K}^2$ / $g_{N\Sigma K}^2$). The SU(6) prediction of
this quantity is 27~\cite{swa63} which would nearly explain the observed
value of R. However, $\pi$-exchange mechanism is important for these
reactions which taken together with the $K$ exchange process leads to a
considerably lower value~\cite{sew99}($\sim$ 3.6) for $R$. Of course, the
$\Sigma^0$ production can be suppressed by the $\Sigma N \to \Lambda N$
conversion process in the $\Sigma-p$ final state interaction. However, there
is no clear evidence in support of the fact that the whole of the observed
enhancement is really due to the produced $\Sigma$ particle being converted
to $\Lambda$ by the FSI effects.

Several theoretical studies have been performed to understand these data.
These include the calculations of the J\"ulich group~\cite{gas00} and 
of Laget~\cite{lag01}, which take into account both $\pi$ and $K$ exchange
mechanisms and include final state interaction (FSI) effects within 
coupled channels approaches, and the resonance models~\cite{tsu97} that
consider the exchange of $\pi$ and other heavier mesons with excitations
of intermediate baryonic resonant states which are coupled to kaon-hyperon
channels. In yet another work~\cite{shy01}, the near threshold hyperon 
production reported in Ref.~\cite{sew99} (for $\epsilon < 13$ MeV), have
been analyzed within an effective Lagrangian model (ELM) which is developed
\cite{shy96,shy98,shy99,shy01,shy03} to investigate the particle production
in nucleon-nucleon ($NN$) collisions.

In the ELM, the initial interaction between two incoming nucleons is
modeled by an effective Lagrangian which is based on the exchange
of the $\pi$, $\rho$, $\omega$, and $\sigma$ mesons. The coupling
constants at the nucleon-nucleon-meson vertices are determined by
directly fitting the T-matrices of the nucleon-nucleon ($NN$) scattering
to the proton-proton and proton-neutron scattering data in the relevant
energy region. In these calculations the pseudovector (PV) coupling is
used for the nucleon-nucleon-pion vertex as it is consistent with the
chiral symmetry requirement of the QCD~\cite{wei68} and also it 
leads to negligible contributions from the negative energy states
("pair suppression phenomena")~\cite{mac87}. The particle production within
the ELM proceeds via excitation of the relevant intermediate baryonic
resonant states.  To describe the near threshold data, the FSI effects in the
final channel are included within the framework of the Watson-Migdal
theory~\cite{wat52,gil64}. ELA has been used so far to describe the
$pp \to pp\pi^0$, $pp \to pn\pi^+$~\cite{shy98,shy96}, $pp \to p K^+Y$
\cite{shy99,shy01} as well as $pp \to pp e^+e^-$~\cite{shy03} reactions.

In Ref.~\cite{shy01}, $N^*$(1650), $N^*$(1710), and $N^*$(1720) baryonic
resonances were included as intermediate states in the ELM calculations of 
total cross sections of both $pp \to p\Lambda K^+$ 
($\sigma_{tot}^{p\Lambda K^+}$) and $pp \to p\Sigma^0 K^+$ 
($\sigma_{tot}^{p\Sigma^0 K^+}$) reactions. In that study, it was concluded
that the contributions of the $N^*(1650)$ resonance state dominate both these
reactions for $\epsilon$ values below 13 MeV. It was also noted that
only with contributions of this resonance state included in both
$\sigma_{tot}^{p\Lambda K^+}$ and $\sigma_{tot}^{p\Sigma^0 K^+}$, can the 
theory explain (within a factor of 2) the large values of $R$ in this 
very close-to-threshold energy regime.

Recently, the COSY-11 collaboration has published~\cite{kow04} 
data for both total cross sections for these reactions as well as their
ratios in an extended energy regime (corresponding to $\epsilon$ values up
to 60 MeV). It is observed that $R$ decreases strongly for $\epsilon$ between
10 MeV - 20 MeV. Beyond this the slope of $R$ is much smaller and the
ratio slowly approaches to values observed at higher energies.
Very recently, both COSY-11 and COSY-TOF collaborations have announced
measurements of $pp \to p\Sigma^+ K^0$ and $pp \to n\Sigma^+ K^+$
reactions~\cite{roz05,bus05}. The analyzes of the data are in progress
and the numerical values for the corresponding cross sections
are likely to be available soon. Furthermore, there are also proposals to
measure the associated hyperon production in $pn$ collisions. Already the
ANKE collaboration has reported \cite{bus04} the estimated ratio of the cross
sections for $pn \to n\Lambda K^+$ and $pp \to p\Lambda K^+$ reactions at 
two near threshold beam energies.

The aim of this paper is to investigate how far can the trends of the data 
in the larger regime of near threshold energies be explained within the ELM.
We would like to know if there are changes in the pattern of the relative
contributions of various resonances to the total cross sections of the two
reactions for $\epsilon$ values larger than 13 MeV. Furthermore, in view of
the experimental activity on the $pp \to p\Sigma^+ K^0$ and 
$pp \to n\Sigma^+ K^+$ reactions, it would be useful to give the predictions
of our model for these channels. We also give result for the ratio of the
$pp \to p\Lambda K^+$ and $pn \to n\Lambda K^+$ reactions at a near
threshold beam energy. 

In the present form of our effective Lagrangian model, the 
energy dependence of the cross section due to FSI is separated from
that of the primary production amplitude and the total amplitude is
written as~\cite{wat52}
\begin{eqnarray}
A_{fi} & = & M_{fi}(pp \rightarrow pYK^+) \cdot T_{ff},
\end{eqnarray}
where $M_{fi}(pp \rightarrow pYK^+)$ is the primary associated 
$YK$ production amplitude, while $T_{ff}$ describes the rescattering
among the final particles which goes to unity in the limit of no FSI.
The Coulomb effects were included by using the modified formula for 
the effective range expansion of the phase-shift as discussed in~\cite{shy99}. 
This type of approach has been used extensively to describe the
pion~\cite{dub86,mei98}, $\eta$-meson~\cite{mol96,dru97,del04},
associated hyperon~\cite{sib00}, and $\phi$-meson \cite{tit00} production
in $pp$ collisions.

The amplitude $M_{fi}$ is determined by following the procedure described
in Ref.~\cite{shy99}. The associated $K^+ Y$ production proceeds via the
excitation, propagation and decay of $N^*(1650)(\frac{1}{2}^{-})$, 
$N^*(1710)(\frac{1}{2}^{+})$, and $N^*(1720)(\frac{3}{2}^{+})$ intermediate
resonant states as below 2 GeV center of mass (c.m.) energy, only these
resonance have significant decay branching ratios into the $KY$ channels. 
Since all the three resonances can couple to the meson-nucleon channels 
mentioned earlier, we require the effective Lagrangians
for all the four resonance-nucleon-meson vertices corresponding to
all the included resonances. One has the freedom of choosing either
psuedoscalar (PS) and PV couplings for the $N^*N\pi$ and $N^*\Lambda Y$
vertices.  In Ref.~\cite{shy99}, $\sigma_{tot}^{p\Lambda K^+}$ was 
calculated using both PS and PV couplings for these vertices. However, it
was noted that the PS couplings for these vertices are clearly favored by
the data. In this study we have used the PS couplings for these vertices.
For all the details about the effective Lagrangians for these vertices and
the determination of the coupling constants appearing therein, the 
propagators for the exchanged mesons and intermediate resonances and
expressions for various amplitudes involved in $M_{fi}$, we refer to
Ref.~\cite{shy99}. The amplitude $T_{ff}$ has been calculated by
following the Jost function~\cite{wat52,gil64} method using the Coulomb
modified effective range expansion of the phase-shift in the same way as
discussed in Refs.~\cite{shy98,shy99,shy01}. The required effective range
and scattering length parameters were the same as those used in
Refs.~\cite{shy99,shy01} (set $\tilde{A}$ of Ref.~\cite{reu94}).

The factorization of the total amplitude into those of the primary production
and FSI [Eq.~(1)], enables one to pursue the diagrammatic approach for the 
meson production process within an effective Lagrangian model and investigate
the role of various meson exchanges and resonances. At the same time it
also allows the inclusion FSI effects among all the three outgoing particles.
However, some difficulties inherent in the application of the Watson-Migdal
method in calculations of the total amplitude for such reactions,
should be kept in mind. Firstly, it has been argued in Refs.~\cite{han99,bar01}
that although the energy dependence of the production process may be
described correctly by Eq.~(1) (particularly for the production of heavier
mesons), its absolute magnitude could be uncertain because of the 
off-shell effects. Secondly, in Ref.~\cite{gas05} it is shown that the  
application of the Jost function method (with effective range approximation)
for extracting the scattering length ($a_S$) parameter for the $\Lambda - p$
final state interaction from the $pp \to p\Lambda K^+$ data, leads to larger
values of $a_S$ as compared to those predicted by $YN$ interaction models.
Thirdly, the Watson-Migdal approximation has no scope for the inclusion of
the coupled channels effects.  

The off-shell effects can be included by multiplying the on-shell FSI 
amplitude with a form factor as is done in Refs.~\cite{lag01,gas00,shy01}.
We have followed here exactly the same procedure as in Ref.~\cite{shy01}.
It should be noted that in this method, both absolute magnitude as well as
shapes of the FSI factor are affected by the off-shell corrections. This 
form factor approach  can be improved by using the off-shell
structure of some realistic $YN$ interaction which may take care, to some
extent, of the second criticism described above. However, it could imply going
beyond the factorization approach of Eq.~(1). A full solution of the problem 
clearly has to wait until a fully relativistic theory of the two-body $\to$
three-body reactions is developed where final state interactions between
outgoing particles are dynamically incorporated in the production amplitude
(perhaps by including diagrams with interaction lines between outgoing 
particles).   
\begin{table}[here]
\begin{center}
\caption {Resonance parameters, the branching ratios for their decay to 
various channels and the corresponding coupling constants obtained therefrom. 
Coupling constants at the $N^*N\omega$ vertices are obtained from the vector
meson dominance hypothesis (see, {\it e.g.} Ref.~\cite{shy99}).}
\vspace{0.5cm}
\begin{tabular}{|c|c|c|c|c|}
\hline
Resonance  & Width & Decay channel & Branching & $g$ \\
           & (GeV) &                & ratio   &    \\
\hline
$N^*$(1710)& 0.100 & $N\pi$      & 0.150     & 1.0414  \\
           &       & $N\rho$     & 0.150     & 4.1421  \\
           &       & $N\omega$   &           & 1.2224  \\
           &       & $N\sigma$   & 0.170     & 0.6737  \\
           &       & $\Lambda K$ & 0.150     & 6.1155  \\
           &       & $\Sigma K$  & 0.030{\footnote{value suggested 
in~\cite{pdg04,pen02} 0.07$\pm$0.06}}
     & 7.4393  \\  
$N^*$(1720)& 0.150 & $N\pi$      & 0.150     & 0.1469  \\
           &       & $N\rho$     & 0.700     & 19.483  \\
           &       & $N\omega$   &           & 16.766  \\
           &       & $N\sigma$   & 0.120     & 1.5557  \\
           &       & $\Lambda K$ & 0.080     & 1.0132  \\
           &       & $\Sigma K$  & 0.060{\footnote{value suggested 
in~\cite{pdg04,pen02} 0.09$\pm$0.03}}
          & 3.0651  \\  
$N^*$(1650)& 0.150 & $N\pi$      & 0.700     & 0.8096 \\
           &       & $N\rho$     & 0.08      & 2.6163 \\
           &       & $N\omega$   &           & 1.8013 \\
           &       & $N\sigma$   & 0.025     & 2.5032  \\
           &       & $\Lambda K$ & 0.070     & 0.7658  \\
           &       & $\Sigma K$  &           & 0.4500  \\  
\hline 
\end{tabular}
\end{center}
\end{table}

Both magnitude and sign of the coupling constant $g_{N^*_{1/2^-}\Sigma K}$
of the $N^*(1650)\Sigma K$ vertex, are uncertain. This is also, to some
extent, true for those of the $N^*(1710)\Sigma K$ and $N^*(1720)\Sigma K$
vertices because the corresponding decay branching ratios 
are known only within a very broad range of values \cite{pdg04}. In 
Ref.~\cite{shy01}, the magnitude of $g_{N^*_{1/2^-}\Sigma K}$ was taken to 
be 0.45 while its sign was assigned to be negative. These were determined
in a preliminary investigation~\cite{wal00,shy01} where fits were performed
to the available data on the $\pi^+p \to \Sigma^+ K^+$, 
$\pi^-p \to \Sigma^0 K^0$, and $\pi^-p \to \Sigma^-K^+$ reactions in an 
effective Lagrangian coupled channels approach~\cite{feu98}. Subsequently, 
in a more rigorous study \cite{pen02}, all the available data for 
transitions from $\pi N$ to five meson-baryon final states, $\pi N$, 
$\pi \pi N$, $\eta N$, $K\Lambda$, and $K\Sigma$ for center of mass energies 
ranging from threshold to 2 GeV, were fitted simultaneously. In 
these analyzes all the baryonic resonances with spin $\leq \frac{3}{2}$
(extended recently~\cite{shk05} to include also spin $\frac{5}{2}$ resonances)
up to excitation energies of 2 GeV, are included as intermediate states.
These studies indicate that the signs of the coupling constants for the 
vertices $N^*(1650)\Sigma K$, $N^*(1710)\Sigma K$, and $N^*(1720)\Sigma K$
could be identical (see, also, Ref.~\cite{pent02}). We found that a better
fit to the hyperon production data is obtained with a positive sign for
$g_{N^*_{1/2^-}\Sigma K}$ (the same as those of other two $N^*\Sigma K$ 
vertices). Thus while retaining the same value for its magnitude, 
we have used a positive sign for this coupling. Furthermore, the magnitudes
of the coupling constants $g_{N^*_{1/2^+}\Sigma K}$ and 
$g_{N^*_{3/2^+}\Sigma K}$ are updated to values which are more in line with
the recommendations of Refs.~\cite{pdg04,pen02,pent02}.
In Table I, we have shown the resonance properties and the branching ratios for
their decay into various channels adopted for calculating the coupling 
constants of the corresponding vertices.  
\begin{figure}
\begin{center}
\includegraphics[width=1.00 \textwidth]{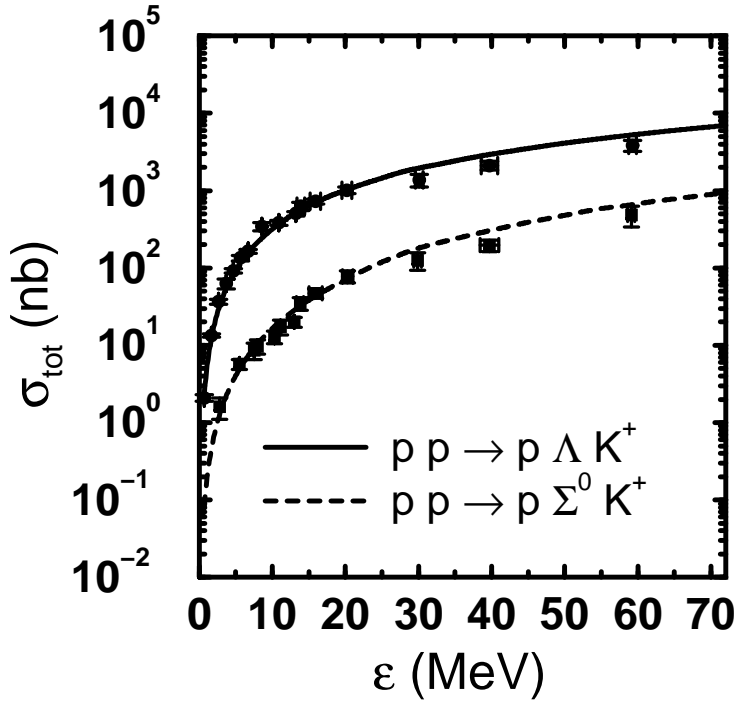}
\vskip -0.3in
\caption{
Comparison of the calculated total cross sections for the
$pp \to p\Lambda K^+$ (solid line) and $pp \to p\Sigma^0K^+$ (dashed line) 
reactions with the corresponding experimental data which are taken from
Refs.~\protect\cite{sew99,kow04}
}
\end{center}
\end{figure}

In Fig~1, we show the total cross sections for $pp \to p\Lambda K^+$
($\sigma_{tot}^{p\Lambda K^+}$) and $pp \to p\Sigma K^+$
($\sigma_{tot}^{p\Sigma^0K^+}$) reactions as a function of the excess energy. 
We note that the calculations which are the coherent sum of all the resonance
excitations and meson exchange processes as described earlier, are in good 
agreement with the data of the COSY-11 collaboration in the entire energy range.
FSI effects are important for both the cases, even though the $p-\Sigma^0$
FSI is somewhat weaker than the $p-\Lambda$ one. The extended energy domain
data of COSY-11 collaboration do indeed put a constraint on the $Y$-nucleon
interaction as it was noted that the data in the entire energy 
range can be satisfactorily explained only with set ($\tilde{A}$) out of
several sets of effective range and scattering length parameters for the 
hyperon-nucleon interaction given in Ref.~\cite{reu94}.  
\begin{figure}
\begin{center}
\includegraphics[width=0.95 \textwidth]{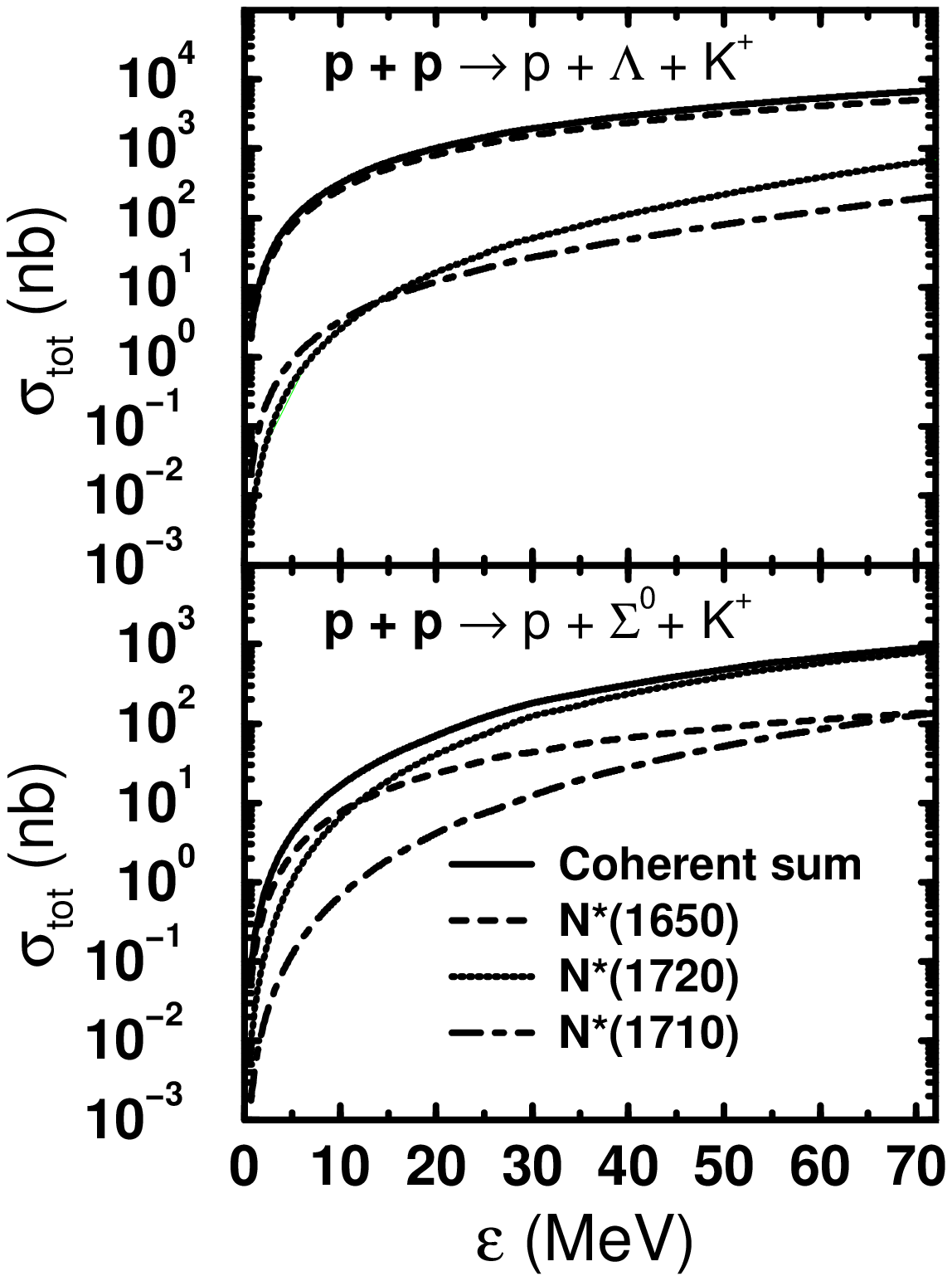}
\vskip -0.1in
\caption{
Contributions of $N^*(1650)$ (dashed line), $N^*(1710)$
(dashed-dotted) and $N^*(1720)$ (dotted line)
baryonic resonances to the total cross section for $pp \to p\Lambda K^+$
(upper panel) and $pp \to p \Sigma^0 K^+$ (lower panel) reactions. 
Their coherent sum is shown by solid lines.
}
\end{center}
\end{figure}

The individual contributions of various nucleon resonances to 
$\sigma_{tot}^{p\Lambda K^+}$ and $\sigma_{tot}^{p\Sigma^0K^+}$
are shown in Fig.~2. We note that in both the cases, the cross sections
are dominated by the contributions from the $N^*(1650)$ resonance excitation
for $\epsilon \leq 13$ MeV which was the energy range of the earlier 
COSY-11 data~\cite{sew99}. This is in agreement with the observations
made in Ref.~\cite{shy01}. Since $N^*(1650)$ is the lowest energy baryonic
resonance which can decay to $YK^+$ channels, its dominance is to be expected
in these reactions at very close-to-threshold beam energies. 
\cite{fal97}. In this energy regime the relative
dominance of various resonances is determined by the dynamics of the
reaction where a difference of about 60 MeV in excitation energies of
$N^*(1650)$ and $N^*(1710)$ resonances plays a crucial role.
 
However, for $\epsilon$ values beyond 15 MeV, while the
$pp \to p\Lambda K^+$ reaction continues to be dominated by
the $N^*(1650)$ excitation, the $pp \to p\Sigma^0K^+$ reaction gets
significant contributions also from higher mass resonances. In fact,
$N^*(1720)$ resonance, dominates this channel for $\epsilon > 30$ MeV.
Simple kinematics suggests that for a resonance intermediate state to 
contribute significantly to the cross section, the values of $\epsilon$ 
should be  $< Q [ = m_{N^*} + \Gamma_{N^*}/2 - m_Y - m_K$, where $m_{N^*}$ 
and $\Gamma_{N^*}$ are the mass and the width of the resonance, respectively].
For $N^*(1650)$, the values of Q are 38 MeV and 115 MeV for
$pp \to p\Sigma^0 K^+$ and $pp \to p\Lambda K^+$ reactions, respectively.
Therefore, the range of $\epsilon$ in which this resonance should be 
contributing significantly is expected to be quite narrow in the case of the 
$pp \to p\Sigma^0K^+$ reaction as compared to the $pp \to p\Lambda K^+$ case.
In this context, it would be worthwhile to note that in the recent 
Dalitz plot analyzes of data for the $pp \to p\Lambda K^+$ reaction by the   
COSY-TOF collaboration (see, eg., Refs.~\cite{eyr03,bus05}), it has been 
observed that even at the beam momentum of 2.85 GeV/c ($\epsilon
\approx$ 171 MeV), this reaction is dominated by the contributions from the
$N^*(1650)$ resonance excitation.  Similar analysis of the
$pp \to p\Sigma^0K^+$ reaction would be very instructive and useful.

Since in some of the earlier resonance model calculations the excitation of
the $N^*(1650)$ resonance intermediate state was not included in the calculation
of $\sigma_{tot}^{p\Sigma^0K^+}$, we emphasize the importance of this resonance
in describing the $pp \to p\Sigma^0K^+$ at very close-to-threshold energies in
Fig.~3. In this figure we we have shown $\sigma_{tot}^{p\Sigma^0K^+}$ 
(upper panel) and the ratio ($R$) of $\sigma_{tot}^{p\Lambda K^+}$ and 
$\sigma_{tot}^{p\Sigma^0K^+}$ (lower panel) as a function of $\epsilon$,
with and without including the contributions of $N^*(1650)$ intermediate 
resonance state in $\sigma_{tot}^{p\Sigma^0K^+}$. It is clear that without
contributions of the $N^*(1650)$ resonance, $\sigma_{tot}^{p\Sigma^0K^+}$ is
underpredicted roughly by an order of magnitude for $\epsilon$ values very
close to the production threshold. As $\epsilon$ increases the contributions
of this resonance becomes lesser and lesser important.
\begin{figure}
\begin{center}
\includegraphics[width=0.85 \textwidth]{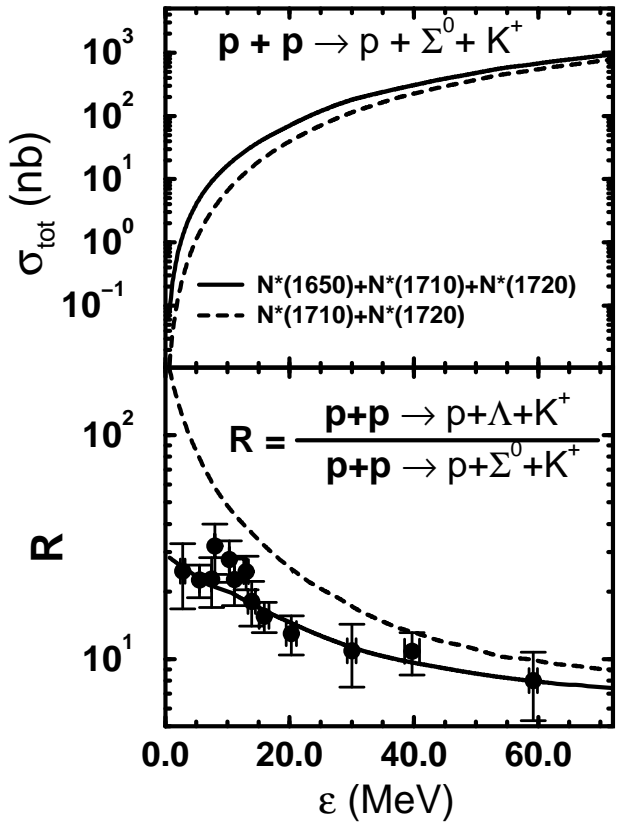}
\vskip -0.1in
\caption{
Role of $N^*(1650)$ resonance in the total cross section for the 
$pp\to p\Sigma^0K^+$ reaction (upper panel) and in the ratio $R$ of
the $pp \to p\Lambda K^+$ and $pp\to p\Sigma^0 K^+$ 
reactions (lower panel) as a function of excess energy. The solid lines
show the results of full calculations while the dashed lines are the ones
obtained without including the $N^*(1650)$ resonance contributions in the
cross sections for the $pp\to p\Sigma^0 K^+$ reaction. In both the
cases, full $\sigma_{tot}^{p\Lambda K^+}$ including contributions from
all the three resonant states, have been used in the calculations o $R$.
The data are from Refs.~\protect\cite{sew99,kow04}.  }
\end{center}
\end{figure}

This trend of $\sigma_{tot}^{p\Sigma^0K^+}$ is directly reflected in
the behavior of $R$. In the lower panel of Fig. 3, we have shown our results
for $R$ with and without considering $N^*(1650)$ excitation in
the calculations of  $\sigma_{tot}^{p\Sigma^0K^+}$. In both the results
shown here, we have used full $\sigma_{tot}^{p\Lambda K^+}$ (including 
contributions of all the three resonances) in the calculations of $R$.
It is clear that for
$\epsilon < 20$ MeV, $R$ is overpredicted by factors ranging from 8 to 4 if
$N^*(1650)$ contributions are not included in $\sigma_{tot}^{p\Sigma^0K^+}$.
Due to reducing significance of these contributions to 
$\sigma_{tot}^{p\Sigma^0K^+}$ with increasing $\epsilon$, the difference
between the two calculations become smaller for bigger values of $\epsilon$. 
At larger beam energies the smaller values of $R$, therefore, are due to
the difference in the resonance contribution pattern. In this regime,
the $N^*(1720)$ and $N^*(1710)$ resonances play a different role for both
these reactions. Therefore, this is yet another example of the sensitivity
of the COSY-11 data to the details of the dynamics of resonance 
contributions. It should also be mentioned here that that our full
calculations are able to explain the trend of the energy dependence of $R$
rather well.

Some authors~\cite{gas00,lag01} have used a different picture to understand
these data. In the calculations reported by the J\"ulich group~\cite{gas00},
the initial $NN$ collisions are modeled in terms of  both $\pi$ and $K$
exchange processes and the FSI effects are included in a coupled
channels approach. They show that while $\Lambda K^+$ production channel
is dominated by the $K$ exchange mechanism, both $\pi$ and $K$ exchange
diagrams contribute with almost equal strength to the $\Sigma^0 K^+$ channel.
With the assumption of a destructive interference between the two amplitudes,
this model is able to explain large values of $R$ for $\epsilon$ corresponding
to beam energies very close to the production threshold. In calculations
reported in Ref.~\cite{lag01} too the relative sign of $K$ and $\pi$ exchange
terms is chosen solely by the criteria of reproducing the experimental data,
although in this work the theory has been applied to describe a wider range
of data including the polarization transfer results of the DISTO
experiment~\cite{bal99} and the missing mass distribution obtained in
the inclusive $K^+$ production measurements performed at
SATURNE~\cite{sie94}. However, the observed energy dependence of $R$ is yet
to be explained within both these calculations. More work on these
models (e.g, including the heavy vector meson exchanges~\cite{kai99} and
finding a way to fixed the relative sign of the $\pi$ and $K$ exchange
amplitudes) would be worth pursuing as it is an interesting alternative
approach to the resonance excitation picture.
\begin{figure}
\includegraphics[width=0.50 \textwidth]{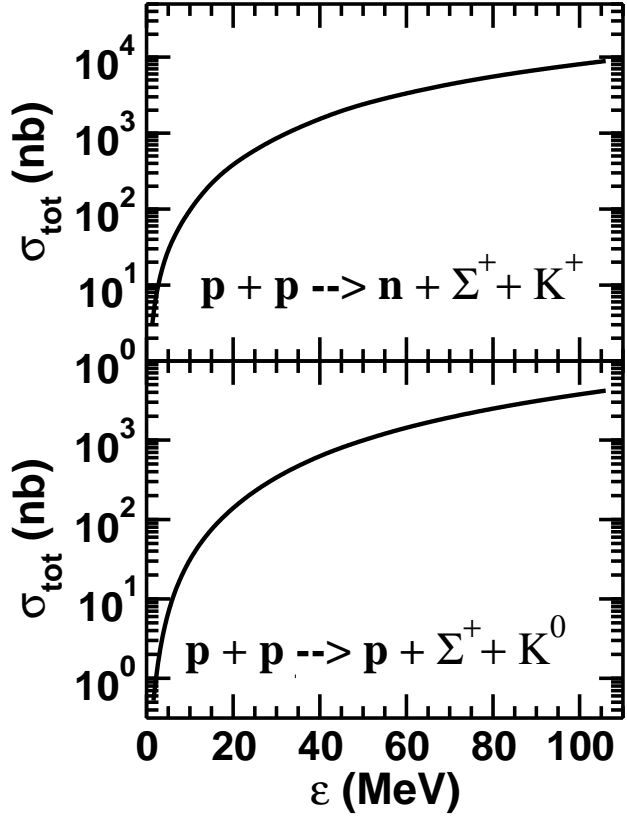}
\caption{
Total cross sections for the $pp \to n\Sigma^+K^+$ and 
$ pp \to p\Sigma^+K^0$ reaction as a function of the 
excess energy.
}
\end{figure}

Recently, both COSY-11~\cite{roz05} and COSY-TOF~\cite{bus05}
collaborations have announced measurements of the total cross sections
for the $pp \to n\Sigma^+K^+$ and $pp \to p\Sigma^+K^0$ reactions. The 
interest in the $\Sigma^+$ production channel stems also from the fact
that it provides a sensitive tool to search for a possible penta quark state
\cite{abd04}. The total cross section data are currently being analyzed.
We show in Fig.~4, the prediction of our model for these reactions.
The isospin factors for different channels are shown in Table II.
Except for slightly different hyperon and strange meson masses, all other
parameters were taken to be the same as those described above. It is hoped
that soon there data would be available to check our predictions.
\begin{table}
\begin{center}
\caption{Isospin factors for various diagrams, Isovector corresponds to
$\pi$ and $\rho$ exchange graphs while isoscalar to $\omega$ and $\sigma$ ones}
\vskip .1in
\begin{ruledtabular}
\begin{tabular}{|ccc|}
graph & isovector & isoscaler\\
\hline
      & {\bf $pn \to n\Lambda K^+$} &   \\
direct   & -1.0 & -1.0 \\
exchange &  2.0 &  0.0 \\

      & {\bf $pp \to p\Sigma^+K^0$} &   \\

direct   & $-\sqrt{2}$ & $-\sqrt{2}$\\
exchange & $-\sqrt{2}$ & $-\sqrt{2}$\\

      & {\bf $pp \to n\Sigma^+K^+$} &    \\

direct   & $2\sqrt{2}$ & 0\\
exchange & $2\sqrt{2}$ & 0\\
\end{tabular}
\end{ruledtabular}
\end{center}
\end{table}

We also calculated the total cross section for the $pn \to n\Lambda K^+$
reaction ($\sigma_{tot}^{n\Lambda K^+}$). Apart from the different isospin
factors, all the parameters were the same in calculations of both  
the reactions. The ratio of $\sigma_{tot}^{n\Lambda K^+}$ and 
$\sigma_{tot}^{p\Lambda K^+}$ is found to be 2.4 at the beam energy of
1.83 GeV. This compares well with the value extracted in Ref.~\cite{bus04}
from the analysis of the inclusive $K^+$ production data.
  
In summary, we have studied the $pp \to p\Lambda K^+$, $pn \to n\Lambda K^+$,
$pp \to p\Sigma^0K^+$, $pp \to n\Sigma^+K^+$, and $pp \to p\Sigma^+K^0$ 
reactions within an effective Lagrangian model in an extended regime of 
near threshold beam energies. The reactions proceed via the excitation of the 
$N^*(1650)$, $N^*(1710)$, and $N^*(1720)$ intermediate baryonic resonant
states.  We confirm that the $N^*(1650)$ resonant state contributes
predominantly to the cross sections of  all these reactions at very
close-to-threshold beam energies. Therefore, in this energy regime, 
hyperon production reactions in nucleon-nucleon collisions provide an
interesting tool for investigating the properties of this negative parity
spin-$\frac{1}{2}$, isospin-$\frac{1}{2}$ resonance much the same way 
way as the $\eta$ meson production probes the lower energy $N^*(1535)$
which has the same parity, spin and isospin.

The inclusion of the contributions of the $N^*(1650)$ resonance in the
cross sections of both $pp \to p\Sigma^0K^+$ and $pp \to p\Lambda K^+$ 
eactions, is essential to explain the experimentally observed large values 
of their ratios at very small values of the excess energies (or beam 
energies). This is also necessary for explaining the beam energy dependence
of this ratio. This result of course assumes that the final state 
interaction effects in the exit channel can be accounted for by the 
Watson-Migdal theory. 

At larger near threshold beam energies (for excess energies between 20-60
MeV), while $pp \to p\Lambda K^+$ reaction continues to be dominated by 
the $N^*(1650)$ resonance, the $pp \to p\Sigma^0K^+$ reaction gets
significant contributions also from the $N^*(1720)$ and $N^*(1710)$
resonances. A Dalitz plot analysis of the data for this reaction would be
very instructive. We also give our predictions for the cross sections
of the $pp \to p\Sigma^+K^0$ and $pp \to n\Sigma^+K^+$ reactions.
It is hoped that data for these channels will soon be available so
that our predictions can be tested. The calculated ratio of the total
cross sections for $pp \to p\Lambda K^+$ and $pn \to n\Lambda K^+$ 
reactions at the beam energy of 1.83 GeV is found to be 2.4 which is very
close to the value extracted in an analysis of the inclusive $K^+$ 
production data.

This work was done when the author was visiting Department of Radiation 
Sciences of the Uppsala University, Sweden. His stay in Uppsala was
supported by the Wenner-Gren Foundation, Stockholm.
He would like to thank Anders Ingemarsson for inviting him to Uppsala and 
for several helpful discussions. Useful conversations  
with Bo H\"oistad and G\"oran F\"aldt are also gratefully
acknowledged.

\end{document}